\documentclass[10pt,conference]{IEEEtran}
\IEEEoverridecommandlockouts
\usepackage{amsmath,amssymb,amsfonts}                           
\usepackage{graphicx}
\usepackage{tikz}
\usepackage{quantikz}
\usepackage{float}
\usepackage{enumitem}

\usepackage[caption=false,font=footnotesize]{subfig} 

\usepackage{orcidlink}
\usepackage{hyperref}                                           
\usepackage[linesnumbered,algoruled,boxed,lined]{algorithm2e}   
\usepackage{tikz}                                               
\usetikzlibrary{quantikz2}                                      
\usepackage{booktabs}                                           
\usepackage{cleveref}                                           
\usepackage{cite} 

\usepackage{etoolbox}
\makeatletter
\patchcmd{\@makecaption}
  {\scshape}
  {}
  {}
  {}
\makeatletter
\patchcmd{\@makecaption}
  {\\}
  {.\ }
  {}
  {}
\makeatother
\def\tablename{Tab.}

\begin{document}
\title{Quantum Optimization Methods for the Generalized Traveling Salesman Problem}

\author{\IEEEauthorblockN{Maximilian Zorn\textsuperscript{\orcidlink{0009-0006-2750-7495}}, Melinda Braun, Michael Ertl, Tommy Kiss, Sara Juarez Oropeza,\\ Claudia Linnhoff-Popien\textsuperscript{\orcidlink{0000-0001-6284-9286}}, and Jonas Stein\textsuperscript{\orcidlink{0000-0001-5727-9151}}}
\IEEEauthorblockA{\textit{Department of Computer Science, \href{https://ror.org/05591te55}{LMU Munich}, Munich, Germany}}}
\maketitle

\begin{abstract}
This paper studies quantum optimization baselines for the Generalized Traveling Salesman Problem (GTSP), a clustered routing problem that naturally models variant selection and sequencing problems under discrete alternatives. We propose a novel GTSP QUBO formulation focused on maintaining feasible solutions for quantum annealing, as well as a hardware-executable gate-based pipeline utilizing the Quantum Approximate Optimization Algorithm (QAOA). We implement a constrained QAOA variant using an XY-mixer, which preserves the stepwise Hamming weight in the ideal circuit model, while feasibility with respect to the full GTSP constraints is tracked explicitly during post-processing. We compare the two quantum optimization paradigms on problem instances from GTSPLIB, an established benchmark dataset, and validate against classical state-of-the-art solvers. To mitigate current quantum hardware size limitations, we further extend a preprocessing method to reduce the node count in instance clusters, constructing new NISQ-friendly instances from reduced subsets. Across all tested instances, quantum solvers often produce competitive solution quality when tested on smaller graphs, but exhibit higher runtimes and a sharp degradation in feasibility and scalability as instance size grows. Our evaluation highlights where quantum optimizers can already succeed and which algorithmic bottlenecks, like sampling rates, runtime issues, and other practical failure modes, remain as open problems.

\end{abstract}

\section{Introduction}
\label{sec:introduction}
The Generalized Traveling Salesman Problem (GTSP) extends the classical Traveling Salesman Problem (TSP) by partitioning nodes into disjoint clusters and requiring a tour that visits \emph{exactly one} representative per cluster while minimizing total travel cost~\cite{GTSP_Def}. The GTSP remains $\mathsf{NP}$-hard and is frequently used to model routing and sequencing problems under discrete alternatives. As a modeling tool, the GTSP is versatile in its application, allowing for broad industrial uses such as flexible manufacturing scheduling~\cite{laporte1996some}, automated storage and retrieval systems~\cite{baniasadi2020transformation}, warehouse order picking with multiple stock locations~\cite{karapetyan2012efficient}, as well as various related routing problems with an additional selection component. Industrial robotics commonly presents a wide range of such problems, where a clustered structure in ordered sequential operations arises naturally in tasks like tool-path navigation and motion planning. For example, a robot may need to execute a sequence of subtasks (e.g., drilling, welding, cutting), where each subtask admits multiple feasible execution variants such as different entry points or inverse-kinematics configurations~\cite{suarezruiz2017robotspfastsolution,Dhami_2023}. 

Recent advances in Quantum Computing~\cite{nielsen_chuang_2010} and Quantum Optimization~\cite{abbas2024challenges} have made promising progress on quantum TSP-style formulations of routing problems, typically through QUBO/Ising encodings for quantum annealing and through QAOA on gate-based hardware~\cite{nusslein2022algorithmic,silva2021mapping,chander2024solving}. While these results provide important foundations for conventional routing problems, comparable end-to-end studies for the GTSP remain much rarer compared to the TSP, with quantum-specific formulations and benchmarks missing almost completely.

Motivated by this gap, we propose a set of quantum optimization baselines for conventional GTSP test instances that can be executed on current quantum hardware and compared directly against strong classical algorithms. For our study, we consider known issues and practical bottlenecks from the quantum TSP literature, including limited qubit budgets, embedding overhead, circuit depth, noise, and the difficulty of maintaining feasible solutions (cf. \cite{silva2021mapping,chander2024solving,padmasola2025solving}). Our evaluation focuses on understanding which formulations are practically viable, where and why they fail under realistic hardware constraints, and how far preprocessing can extend the solvable instance range on today’s Noisy Intermediate-Scale Quantum (NISQ) devices. The contributions of this paper are as follows:

\begin{enumerate}[label=\roman*.]
\item We propose a novel GTSP QUBO encoding with quadratic penalty terms specifically focused on producing feasible solutions. We further provide a hardware-oriented constrained QAOA pipeline based on a one-hot-preserving XY-mixer, and report feasibility-aware decoding, sampling rates, runtime, and practical failure modes of our algorithms.
\item We benchmark both quantum optimization approaches against classical state-of-the-art solvers on the established GTSP benchmark dataset GTSPLIB.
\item We introduce a practical preprocessing method that reduces benchmark-derived GTSP instances to NISQ-feasible sizes and provide empirical insights on scalability, runtime, and practical failure modes.
\end{enumerate}

We begin by summarizing the technical background on GTSP, QUBO/Ising models and quantum optimization in Section~\ref{sec:background}, before reviewing related literature on relevant quantum approaches in Section~\ref{sec:relatedwork}. We detail our QUBO formulation, our QAOA/annealing pipelines and NISQ-feasible instance preprocessing, as well as our experimental setup (Sections~~\ref{sec:methodology}\&\ref{sec:setup}). Sections~\ref{sec:evaluation}\&~\ref{sec:discussion} provide an in-depth discussion and interpretation of solution quality, scalability and feasibility of the evaluation results, after which we conclude our findings and outline directions for future work.

\section{Background}
\label{sec:background}
\subsection{Generalized Traveling Salesman Problem (GTSP)}
The Generalized Traveling Salesman Problem is an optimization problem focused on finding a minimum-cost tour $S^\ast = \arg\min_{S \in \mathcal{S}} \operatorname{cost}(S)$, where $\mathcal{S}$ denotes the set of feasible tours and $\operatorname{cost}(S)$ the total travel cost of a tour. The GTSP is defined on an undirected connected, weighted graph $G = (V,E)$, where $V$ denotes the set of nodes and $E \subseteq V \times V$ the set of edges. Each edge $(i,j) \in E$ is assigned a non-negative weight given by the function $w : E \rightarrow \mathbb{R}_{+}$. The node set $V$ is partitioned into $K$ mutually exclusive subsets $C = \{C_1, C_2, \dots, C_K\}$ for $ C_k \subseteq V$, such that $\bigcup_{k=1}^{K} C_k = V$ and $C_i \cap C_j = \emptyset \quad \forall i \neq j.$ Each subset $C_k$ is referred to as a cluster. The set of feasible tours, that visit each cluster exactly once returning to the origin, is defined as $\mathcal{S}=\{(v_1, \dots, v_K)\;|\;v_k \in C_k,\; k = 1,\dots,K \}$ with the corresponding cyclic tour $(v_1, v_2), (v_2, v_3), \dots, (v_{K-1}, v_K), (v_K, v_1).$ The cost of a tour is defined as $\operatorname{cost}(S) = \sum_{(i,j) \in S} w_{ij}$, where $ w_{ij}$ is the weight of the edge $(v_i, v_j) \in E$.

\subsection{Quadratic Unconstrained Binary Optimization (QUBO)}
\label{sec:background_qubo}
QUBO represents the class of binary optimization problems as the minimization of a quadratic function $f:\left\lbrace 0,1\right\rbrace^n\rightarrow\mathbb{R}$ over binary variables $\arg\min_{x \in \left\lbrace 0,1\right\rbrace^n} x^TQx$, for some QUBO matrix $Q\in\mathbb{R}^{n\times n}$ such that $x^TQx=f(x)$. QUBO matrices are equivalent to Ising Hamiltonians and thus suitable for native encoding on quantum devices. Minimizing the total energy of the system is then equivalent to finding the solution(s) of minimal cost for any given optimization problem~\cite{glover2019tutorialformulatingusingqubo}. Given that QUBO exactly represents the $\mathsf{NP}$-hard problem of quadratic programming, all $\mathsf{NP}$-hard problems can be formulated as a QUBO matrix~\cite{Lucas_2014_Ising_formulations}. The objective function can be further extended with penalty terms $p:\left\lbrace 0,1\right\rbrace^n\rightarrow\mathbb{R}^+_0$ to encode problem specific constraints through simple addition, i.e., $x^TQx\coloneqq f(x)+\lambda p(x)$, where $p(x)=0$ for valid solutions and $p(x)>0$, otherwise. These penalty terms are scaled by a penalty factor $\lambda>0$ that ensures that constraint violations are sufficiently undesirable in the energy landscape.

\subsection{Quantum Annealing (QA)}
Quantum annealing is an optimization paradigm that exploits the quantum physical phenomenon of adiabatic time evolution and can be used to solve optimization problems. This works by encoding the cost function as an Ising Hamiltonian $H_C=-\sum_{ij}J_{ij}\sigma_i^z\sigma_j^z-\sum_ih_i\sigma_i^z$, such that its ground state $\ket{\varphi}$ encodes the optimal solution of the given optimization problem, i.e., $\bra{\varphi}H_C\ket{\varphi}=\min_xf(x)$. Given an easy to prepare ground state of a known Hamiltonian $H_I$ (e.g., $\ket{+}^{\otimes n}$ and corresponding $H_I\coloneqq -\sum_i \sigma_i^x$), we can use the time evolution operator
\begin{equation}
H(t) = \left( 1 - \frac{t}{T} \right) H_I + \frac{t}{T} H_C,
\end{equation}
that interpolates between $H_I$ and $H_C$, to evolve the ground state of $H_I$ into the desired ground state $H_C$, as governed by the adiabatic theorem~\cite{Born1928}.
Even though the duration required for executing this time evolution in an adiabatic manner is typically exponentially large for $\mathsf{NP}$-hard problems~\cite{Lucas_2014_Ising_formulations}, faster time evolutions are practically possible and often yield close approximations to the optimal solutions. 

\subsection{Quantum Approximate Optimization Algorithm (QAOA)}
The Quantum Approximate Optimization Algorithm, introduced by Farhi \textit{et al.}~\cite{farhi2014quantum}, is a variational quantum algorithm designed to solve combinatorial optimization problems. It can be understood as a parameterized Trotterization of the quantum adiabatic algorithm~\cite{farhi2000quantum}, where the continuous adiabatic evolution is discretized into $p$ alternating layers.

\subsubsection{Problem Formulation}

QAOA operates on binary optimization problems of the form
\begin{equation}
    x^* = \arg\min_{x \in \{0,1\}^n} f(x),
\end{equation}
where $f : \{0,1\}^n \to \mathbb{R}$ is the objective function. The algorithm encodes the objective values as energy levels of a \emph{cost Hamiltonian}
\begin{equation}
    H_C = \sum_{x \in \{0,1\}^n} f(x) \ket{x}\bra{x},
\end{equation}
which is diagonal in the computational basis. The goal is to find the ground state of $H_C$, corresponding to the optimal solution $x^*$.

\subsubsection{Circuit Structure and Optimization}

The QAOA circuit is initialized in the equal superposition state $\ket{+}^{\otimes n} = H^{\otimes n}\ket{0}^{\otimes n}$, followed by $p$ alternating layers of two unitary operators:
\begin{equation}
    \ket{\boldsymbol{\beta}, \boldsymbol{\gamma}} = U_M(\beta_p)\, U_C(\gamma_p) \cdots U_M(\beta_1)\, U_C(\gamma_1) \ket{+}^{\otimes n},
\end{equation}
where $\boldsymbol{\beta}$ and $\boldsymbol{\gamma}$ are free parameters to be optimized classically by minimizing $\langle H_C \rangle$.
The \emph{cost unitary} $U_C(\gamma) = e^{-i\gamma H_C}$ encodes the objective function as a phase, while the \emph{mixer unitary} $U_M$ drives transitions between computational basis states and is typically chosen as the standard transverse-field mixer (X-Mixer)
\begin{equation}
    U_M(\beta) = e^{-i\beta H_M} = \bigotimes_{k=1}^{n} R_X^{(k)}(2\beta),
\end{equation}
where $H_M = -\sum_{i=1}^n \sigma_i^x$. As $p \to \infty$, the output state is guaranteed to converge to the optimal solution for suitable parameter choices such as $\beta_i=1-i/p$ and $\gamma_i=i/p$~\cite{farhi2014quantum, sack2021quantum}.

\subsubsection{Handling Constraints}

For constrained optimization problems, a standard approach augments the cost Hamiltonian with a quadratic penalty term $\lambda \cdot p(x)$ that penalizes constraint violations, converting the constrained problem into an unconstrained one~\cite{Lucas_2014_Ising_formulations}. However, this introduces an additional hyperparameter $\lambda$ and can complicate the optimization landscape~\cite{shirai2025compressed}. Alternative approaches include oracle-based methods that encode constraint satisfaction directly into the cost unitary without modifying the underlying energy spectrum~\cite{bucher2025penalty} and constraint-preserving mixer operators~\cite{hadfield2019quantum} that restrict the quantum evolution to the feasible subspace, the latter of which is adopted in the presented experiments.

\section{Related Work}
\label{sec:relatedwork}
Most prior quantum work on routing problems has focused on the TSP rather than the GTSP. On the annealing side, several studies developed QUBO or closely related Ising formulations for Hamiltonian-cycle and TSP variants, showing how routing problems can be mapped to quantum optimization hardware~\cite{nusslein2022algorithmic}. Experimental studies further demonstrate that quantum annealers and related Ising-type machines can produce meaningful results on small instances, but their scalability is strongly limited by embedding overhead and available qubit resources~\cite{silva2021mapping}.

On the gate-based side, QAOA has become a central approach for routing-type optimization. Chander \textit{et al.}~\cite{chander2024solving} study the QAOA for the TSP and show that while variational quantum heuristics are feasible on small instances, constraint enforcement, circuit depth, and qubit scaling quickly become limiting factors. Follow-up work has therefore focused on improved formulations and mixer design. In particular, Qian \textit{et al.}~\cite{qian2023comparative} compare several QAOA variants for TSP and show that problem-specific mixers, such as the XY-mixer, can improve feasibility and performance over standard transverse-field mixers. Padmasola \textit{et al.}~\cite{padmasola2025solving} further highlight the importance of hardware connectivity and noise for solution quality as instance size increases.

A complementary line of work aims to reduce quantum resource requirements through alternative encodings and subspace-reduction techniques, as demonstrated by Madhus \textit{et al.}~\cite{kalleri2025edge}, who propose an edge-based encoding combined with subspace reduction. Similarly, Glos \textit{et al.}~\cite{glos2022space} introduce a higher-order binary optimization formulation that reduces qubit usage, alongside hybrid encodings tailored for QAOA. Their work highlights the fundamental trade-off between qubit count and circuit depth, emphasizing the need to balance these resources in the design of variational quantum algorithms.

Bak\'{o} \textit{et al.}~\cite{bako2025prog} introduce Prog-QAOA, addressing another key challenge in the implementation of quantum algorithms, namely the high entry barrier associated with requiring detailed knowledge of low-level quantum computing concepts to formulate real-world problems as QUBO instances or corresponding Hamiltonians. Prog-QAOA is a framework in which quantum circuits are generated directly from classical pseudocode descriptions, rather than relying on standard QUBO formulations. While this approach enables more structured circuit design and can reduce qubit count, circuit depth, and gate complexity for problems such as Max-K-Cut and the Traveling Salesman Problem, these improvements depend on problem-specific constructions, which may limit generality and hinder direct comparability with existing encoding-based methods.

Taken together, these studies provide the main building blocks for quantum (G)TSP approaches, but evidence for GTSP remains limited compared with the much broader body of work on TSP.

\section{Methodology}
\label{sec:methodology}

In this section, we present our purely quantum and hybrid quantum solver implementations alongside a preprocessing pipeline designed to address scalability challenges in GTSP optimization. To mitigate the limited problem sizes supported by current Noisy Intermediate-Scale Quantum (NISQ) hardware, we introduce a classical preprocessing strategy based on the Nearest Nodes to Clusters approach~\cite{el2021pre} for instance reduction. These techniques enable an evaluation of the maximum instance sizes tractable on current quantum hardware, independent of solution quality.

\subsection{Quadratic-unconstrained-binary-optimization Generalized Traveling Salesman Problem (Q-GTSP)}
\label{subsec:gtsp_qubo}
In this subsection, we explore how the GTSP can be solved using quantum annealing. For this, we construct a suitable QUBO formulation by expanding on well-known QUBO formulations of the standard TSP~\cite{10.1007/978-981-97-7801-0_12}.
\subsubsection{QUBO formulation}
\label{subsec:QUBO_Formulation}
We use a one-hot encoding for the GTSP with $N \cdot K$ many binary decision variables (where $K$ denotes the number of clusters and $N$ the number of nodes). For this, let $x_{c,i} \in \{0,1\}$, where $c \in \{0,\dots,K-1\}$ and $ i \in \{0,\dots,N-1\}$,
such that $x_{c,i}=1$ encodes that node $i$ is visited in $c$-th position.

Let $w_{ij}$ be the edge weight from node $i$ to node $j$.
The objective is to minimize the tour cost over consecutive steps (cyclically):
\begin{equation}
E_{\text{cost}}(x)
=
\sum_{c=0}^{K-1}\sum_{i=0}^{N-1}\sum_{j=0}^{N-1}
w_{ij}\; x_{c,i}\,x_{(c+1)\bmod K,\,j}.
\end{equation}

Furthermore, exactly one node must be visited at each step $c \in \{0, \dots, K-1\}$, which can be described with the penalty term
\begin{equation}
    p_0(x) \coloneqq \lambda \sum_{c=0}^{K-1} \left( \sum_{i=0}^{N-1} x_{i, c} - 1 \right)^2.
\end{equation}

Additionally, for each cluster $C_m$, exactly one node within that cluster is visited throughout the entire sequence, as ensured by the penalty term
\begin{equation}
    p_1(x) \coloneqq \lambda \sum_{m} \left( \sum_{c=0}^{K-1} \sum_{i \in C_m} x_{i, c} - 1 \right)^2.
\end{equation}

Finally, we penalize transitions between nonexistent edges through

\begin{equation}
     p_2(x) \coloneqq \lambda \sum_{(i,j) \in \mathcal{E}_{\textnormal{zero}}} x_{i} x_{j},
\end{equation}
where $\mathcal{E}_ {\textnormal{zero}} \coloneqq \{ (i, j) : D_{i,j} = 0, i \neq j \}$.

\subsubsection{Penalty Weight} \label{Metho:PenaltyTerm}
Let $w_{ij}$ denote the edge weights of the graph and let
\[
w_1 \ge w_2 \ge \dots \ge w_{|E|}
\]
be all edge weights sorted in non-increasing order.
This allows us to straightforwardly define the penalty weight $\lambda$ via
\begin{equation}
\lambda\coloneqq \sum_{k=1}^{K} w_k \;+\; 1,
\end{equation}
which serves as a trivial upper bound for any valid solution cost. As any invalid solution is then penalized by at least this penalty factor, the cheapest invalid solution is provably more expensive than the most expensive valid solution.

\subsection{Constrained QAOA with One-Hot Preserved XY-Mixer (C-QAOA)}
Aiming for a balance between NISQ-friendliness and constraint-satisfaction, our QAOA implementation employs the same basic QUBO formulation that was proposed in
\Cref{subsec:gtsp_qubo}, while enforcing the one-hot constraint represented by $p_0(x)$ using XY-mixers.  
For this, variables belonging to the same step $t$ form a one-hot partition $P_t$. To minimize the required circuit depth, we apply a XY-mixer within each partition using a low-depth ring topology. To further ensure the lightest possible hardware demands for the initial state preparation, we initialize in a single feasible (randomly selected) one-hot computational basis state rather than the theoretically desired uniform superposition over all feasible states. Formally, the mixer Hamiltonian and its corresponding unitary can be written as 
\begin{equation}
    H_M = \sum_t \sum_{(u,v)\in E_t} (X_uX_v + Y_uY_v) 
\end{equation} 
and $U_M(\beta)=e^{-i\beta H_M}$, where $E_t$ denotes the edges within partition $P_t$. 
Note, however, that only one of the three constraints is addressed with this circuit depth-minimal construction. We still need to use the penalty terms $p_1(x)$ and $p_2(x)$ for the other constraints, as no short circuit depth mixer construction is known that could also address these in combination with $p_0(x)$.



\subsection{Instance Preprocessing for Quantum Solvers: Extended Nearest Nodes to Clusters }
Since one of the main limitations of our solvers is the restricted number of available qubits, we apply a preprocessing technique to reduce the size of the GTSP instances before solving them on quantum hardware. This preprocessing is based on the Nearest Nodes to Clusters (NN2C) approach~\cite{el2021pre}, which was originally proposed for symmetric GTSP instances. The original NN2C algorithm selects, for each cluster, nodes that are closest to other clusters and removes the other nodes, since they are less likely to be part of a good tour, thereby reducing the instance dimension.

\begin{figure}[ht!]
  \centering
  \graphicspath{{tikz/}}
  \def\svgwidth{0.7\linewidth}
  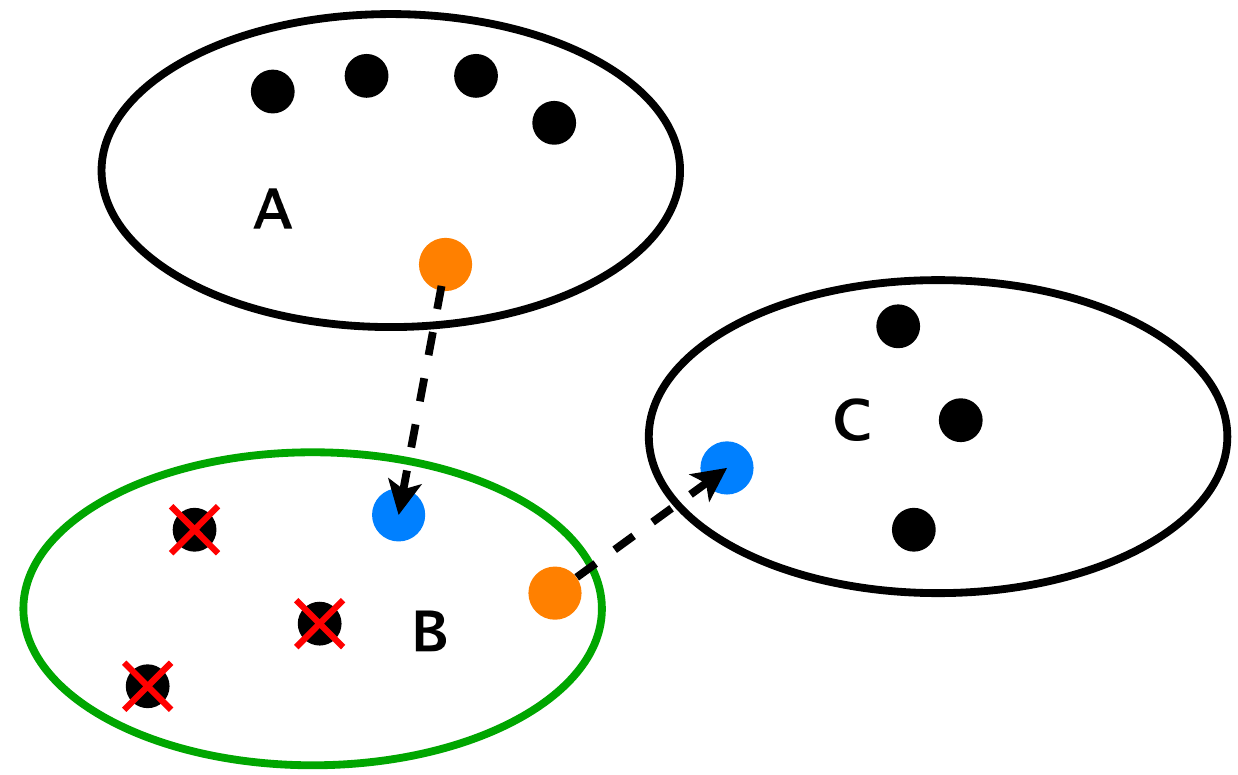
  \caption{Example illustrating the application of the NN2C extended preprocessing method: Here, for cluster B, the best entry node (blue) and best exit node (orange) are selected. All not selected nodes (red X) are removed. This step is repeated for all clusters.}
  \label{fig:NN2C_visual}
\end{figure}

To address the challenges posed by asymmetric GTSP instances in the GTSPLIB dataset~\cite{GTSPLIB}, we extend the NN2C approach by explicitly retaining at least one entry node and one exit node per cluster. Entry nodes are defined as nodes that minimize incoming distances from other clusters, while exit nodes minimize outgoing distances to other clusters. In cases where multiple nodes share the same minimal cost, a secondary selection criterion based on the average cost in the opposite direction is applied. This ensures a deterministic and stable node selection.
In our extended version of the NN2C algorithm (see \Cref{algo:NN2C}), we iterate over all clusters and evaluate all nodes within each cluster. For every node, we compute the minimum distance to any other node in a different cluster. Among all nodes in the cluster, we retain the node with the smallest such incoming cost as the best entry node. Analogously, we identify the best exit node by selecting the node with the minimum outgoing distance to any other cluster. This procedure is illustrated in \Cref{fig:NN2C_visual}.

If multiple nodes have identical minimal incoming costs, we compare their overall outgoing costs and select the node with the smaller average outgoing distance, as this node is potentially more suitable in a global tour. Conversely, if multiple nodes have identical minimal outgoing costs, the node with the smaller average incoming distance is selected. As a result, for each cluster, exactly one best entry node and one best exit node are chosen. The same node may fulfill both roles. Each cluster is reduced in nodes, and a new GTSP instance is constructed from the resulting reduced clusters.
The resulting reduced instances contain at most twice as many nodes as clusters and, in the best case, reduce the GTSP to a TSP if each cluster is represented by a single node. Despite this reduction, the generated instances still represent complete tours, similar to the original GTSPLIB instances, and preserve several of their structural characteristics. 

We note as limitation to this approach, that there is no guarantee that the optimal tour of the original GTSPLIB instance is preserved in the reduced instance. However, the substantially reduced problem size enables these instances to be solved using quantum solvers. Moreover, this preprocessing makes it possible to study GTSPLIB instances that would otherwise be too large for current quantum hardware. The reduced instances are therefore not intended to approximate optimal solutions of the original problem, but to enable feasibility studies and comparative evaluations on quantum hardware.
\begin{algorithm}[h]
\KwIn{GTSP instance $I = (V, \mathcal{C}, w)$}
\KwOut{Reduced instance $I' = (V', \mathcal{C}', w')$}
\For{each cluster $C_i \in \mathcal{C}$}{
    \For{each node $v \in C_i$}{
        compute \textsc{bestIN(v)}\\
        compute \textsc{bestOUT(v)}\\
    }
    break ties in \textsc{bestIN, bestOUT} and form $C_i'$ from resulting nodes\
}
\KwRet{$I'$}
\caption{NN2C extended}\label{algo:NN2C}
\end{algorithm}

\section{Experimental Setup}
\label{sec:setup}


\subsection{Optimizing Variational- and Hyperparameters}

\paragraph{Quantum Annealing Approach (Q-GTSP)}
The Quantum Annealing approach of this work uses the default parameters provided by D-Wave \cite{dwave_quantum_annealing_intro} except for the $num\_reads$ parameter which is set to 1500. This ensures that the probability distribution of the resulting quantum system is thoroughly sampled. Minor embedding is handled automatically by the function \texttt{EmbeddingComposite}.

\paragraph{Gate-Based Approach (C-QAOA)}

Aiming for a particularly NISQ-friendly approach, the variational parameters $(\gamma,\beta)$ are optimized using grid search. This circumvents the computation of exact gradients, which can be very expensive on quantum hardware, since the number of shots scales quadratically with the desired maximal error.
We evaluate a $10\times10$ grid over
$\gamma \in [0.05,\pi]$ and $\beta \in [0.05,\pi/2]$,
using a batch size of one and 1500 shots per circuit evaluation.
Each C-QAOA run is limited to a timeout of 300 seconds per instance.
These parameters were concluded after testing the algorithms with different configurations. As well as cross-checking other works.

\subsection{QPU Usage}

The quantum annealing runs were executed on D‑Wave Systems' Advantage2 QPU, which features a Zephyr topology with over 4{,}400 superconducting qubits and enhanced connectivity~\cite{dwave_advantage_quantum_computer_2026}. 
QAOA experiments were implemented using the Aer simulator through the Qiskit interface, as our available QPU time on IBM Quantum hardware was insufficient for a full experimental evaluation including parameter optimization. The target devices would have been from the Heron processor family, which offer around 156 qubits in a heavy‑hex lattice, with multiple revisions improving coherence and error rates~\cite{ibm_quantum_processor_types_2026}. However, initial tests on IBM QPUs confirmed that the algorithm is in principle compatible with real quantum hardware, and a full QPU-based evaluation remains a natural direction for future work.

\subsection{Dataset}
\label{subsec:Dataset}

To ensure representative benchmarking on reasonably hard problem instances, we evaluated our approaches on established benchmark instances from GTSPLIB~\cite{GTSPLIB}. These instances provide realistic cluster structures and cost patterns. They are widely used in literature and therefore enable meaningful comparisons with existing methods and expedite reproducibility.

Because of the limitations of current quantum hardware, we were only able to solve relatively small instances. In our experiments, the purely quantum solvers can handle instances with at most 25 nodes. 
This strongly restricts the number of GTSPLIB instances that can be solved directly. To expand on the feasibly usable GTSPLIB instances, we apply the following cluster subset sampling strategy to obtain smaller instances that remain compatible with quantum solvers.

\subsection{Cluster Subset Sampling}
\label{subsec:subset_sampling}
To generate smaller GTSP instances from large GTSPLIB benchmarks, we use a single-stage cluster sampling strategy analogous to~\cite{bun2020controlling}. We iteratively select entire clusters at random until a target number of nodes is approximately reached. From the selected clusters, we then construct a new GTSP instance, in which all of the other clusters are simply discarded completely. This method, which we refer to as \textit{Cluster Subset Sampling}, preserves the cluster structure of the GTSP and enables the creation of various benchmark sub-instances with controllable sizes. Even though the resulting instances are reduced in size, they can still retain characteristics of realistic GTSP problems (most importantly the number of nodes per cluster and the pairwise connections between the clusters) while being compatible with the strict limitations of current quantum hardware.
We run our experiments on two datasets generated via subset sampling: Subsample Small (3--5 nodes; \Cref{tab:instance-overview-subsample-3-5-nodes}) and Subsample Medium (3--20 nodes; \Cref{tab:instance-overview-subsample-3-20-nodes}).

\begin{table}[htbp]
  \centering
  \caption{Dimension, original dimension, number of clusters, and qubits for each instance in the experiment group \textit{Subsample Small}. }
  \label{tab:instance-overview-subsample-3-5-nodes}
  \begin{tabular}{lccc}
    \toprule
    Instance & Dimension & Clusters & Qubits\\
    \midrule
    12ftv55\_nodes\_3 & 3 & 2 & 6 \\
    16pr76\_nodes\_3 & 3 & 2 & 6 \\
    6fri26\_nodes\_3 & 3 & 2 & 6 \\
    16eil76\_nodes\_4 & 4 & 2 & 8 \\
    4ulysses16\_nodes\_4 & 4 & 2 & 8 \\
    5ulysses22\_nodes\_4 & 4 & 3 & 12 \\
    6fri26\_nodes\_4 & 4 & 3 & 12 \\
    20gr96\_nodes\_5 & 5 & 4 & 20 \\
    9ftv44\_nodes\_5 & 5 & 2 & 10 \\
    9p43\_nodes\_5 & 5 & 3 & 15 \\
    \bottomrule
  \end{tabular}
\end{table}

\begin{table}[htbp]
  \centering
  \caption{Dimension, original dimension, number of clusters, and qubits for each instance in the experiment group \textit{Subsample Medium}.}
  \label{tab:instance-overview-subsample-3-20-nodes}
  \begin{tabular}{lccc}
    \toprule
    Instance & Dimension & Clusters & Qubits \\
    \midrule
    5ulysses22\_nodes\_3 & 3 & 3 & 9 \\
    9p43\_nodes\_5 & 5 & 3 & 15 \\
    10att48\_nodes\_7 & 7 & 3 & 21 \\
    10hk48\_nodes\_10 & 10 & 3 & 30 \\
    14st70\_nodes\_11 & 11 & 2 & 22 \\
    11ft53\_nodes\_13 & 13 & 4 & 52 \\
    20kroD100\_nodes\_15 & 15 & 4 & 60 \\
    20gr96\_nodes\_16 & 16 & 7 & 112 \\
    12brazil58\_nodes\_18 & 18 & 6 & 108 \\
    20rd100\_nodes\_20 & 20 & 7 & 140 \\
    \bottomrule
  \end{tabular}
\end{table}

In addition, we conduct experiments on symmetric and asymmetric original GTSPLIB datasets. For these experiments, all solvers receive original GTSPLIB instances, for which the quantum solvers employ the NN2C extended preprocessing described in \Cref{sec:methodology}. The resulting instances can be grouped into two categories: Preprocess Small (original instances with 14--22 nodes, reduced to 3--5 nodes; \Cref{tab:instance-overview-preprocess-3-5-nodes}) and Preprocess Medium (original instances with 14--100 nodes, reduced to 3--20 nodes; \Cref{tab:instance-overview-preprocess-3-20-nodes}).

\begin{table}[htbp]
  \centering
  \caption{Dimension, original (OG) dimension, number of clusters, and qubits for each instance in the experiment group \textit{Preprocess Small}.}
  \label{tab:instance-overview-preprocess-3-5-nodes}
  \begin{tabular}{lcccc}
    \toprule
    Instance & Dimension & OG Dimension & Clusters & Qubits\\
    \midrule
    3burma14 & 3 & 14 & 3 & 9 \\
    4br17 & 4 & 17 & 4 & 16 \\
    4gr17 & 4 & 17 & 4 & 16 \\
    4ulysses16 & 4 & 16 & 4 & 16 \\
    5gr21 & 5 & 21 & 5 & 25 \\
    5gr24 & 5 & 24 & 5 & 25 \\
    5ulysses22 & 5 & 22 & 5 & 25 \\
    \bottomrule
  \end{tabular}
\end{table}

\begin{table}[h!]
  \centering
  \caption{Dimension, original (OG) dimension, number of clusters, and qubits for each instance in the experiment group \textit{Preprocess Medium}. }
  \label{tab:instance-overview-preprocess-3-20-nodes}
  \begin{tabular}{lcccc}
    \toprule
    Instance & Dimension & OG Dimension & Clusters & Qubits\\
    \midrule
    3burma14 & 3 & 14 & 3 & 9 \\
    6bayg29 & 6 & 29 & 6 & 36 \\
    7ftv33 & 12 & 34 & 7 & 84 \\
    8ftv38 & 13 & 39 & 8 & 104 \\
    14st70 & 14 & 70 & 14 & 196 \\
    10ftv47 & 15 & 48 & 10 & 150 \\
    9ftv44 & 15 & 45 & 9 & 108 \\
    16pr76 & 16 & 76 & 16 & 256 \\
    12ftv55 & 20 & 56 & 12 & 240 \\
    20kroA100 & 20 & 100 & 20 & 400 \\
    \bottomrule
  \end{tabular}
\end{table}

\subsection{Classical Baseline}
\label{subsec:baselines}
For a meaningful baseline, we utilize the classical state-of-the-art algorithm for the GTSP: GLNS, which is an adaptive large neighborhood search algorithm as presented in~\cite{SMITH20171}. For the problem sizes explored, we could verify that GLNS (also called CGLNS later on, to stress that this is a classical method) always found the global optimum using the industrial standard optimization software Gurobi~\cite{gurobi}. 

\section{Evaluation}
\label{sec:evaluation}

\begin{figure}
    \centering
    \includegraphics[width=1\columnwidth]{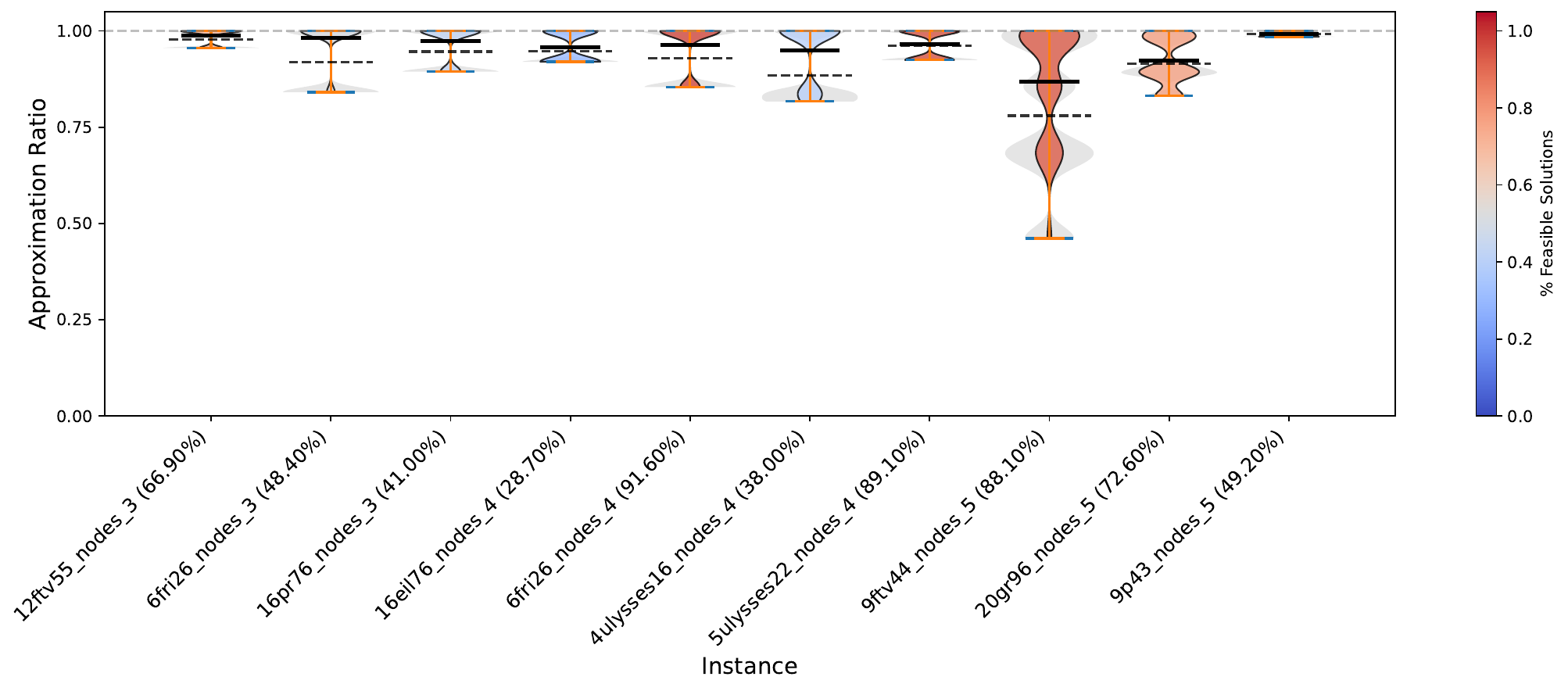}
    \caption{Distribution of solution quality achieved by Q-GTSP on the Subsample Small dataset with 3 to 5 nodes. The violins show the approximation ratio distribution over sampled feasible solutions, while the instance labels report the feasible-shot rate.}
    \label{fig:subsampling_3-5-QGTSP-approx-violin}
\end{figure}

\begin{figure}
    \centering
    \includegraphics[width=1\columnwidth]{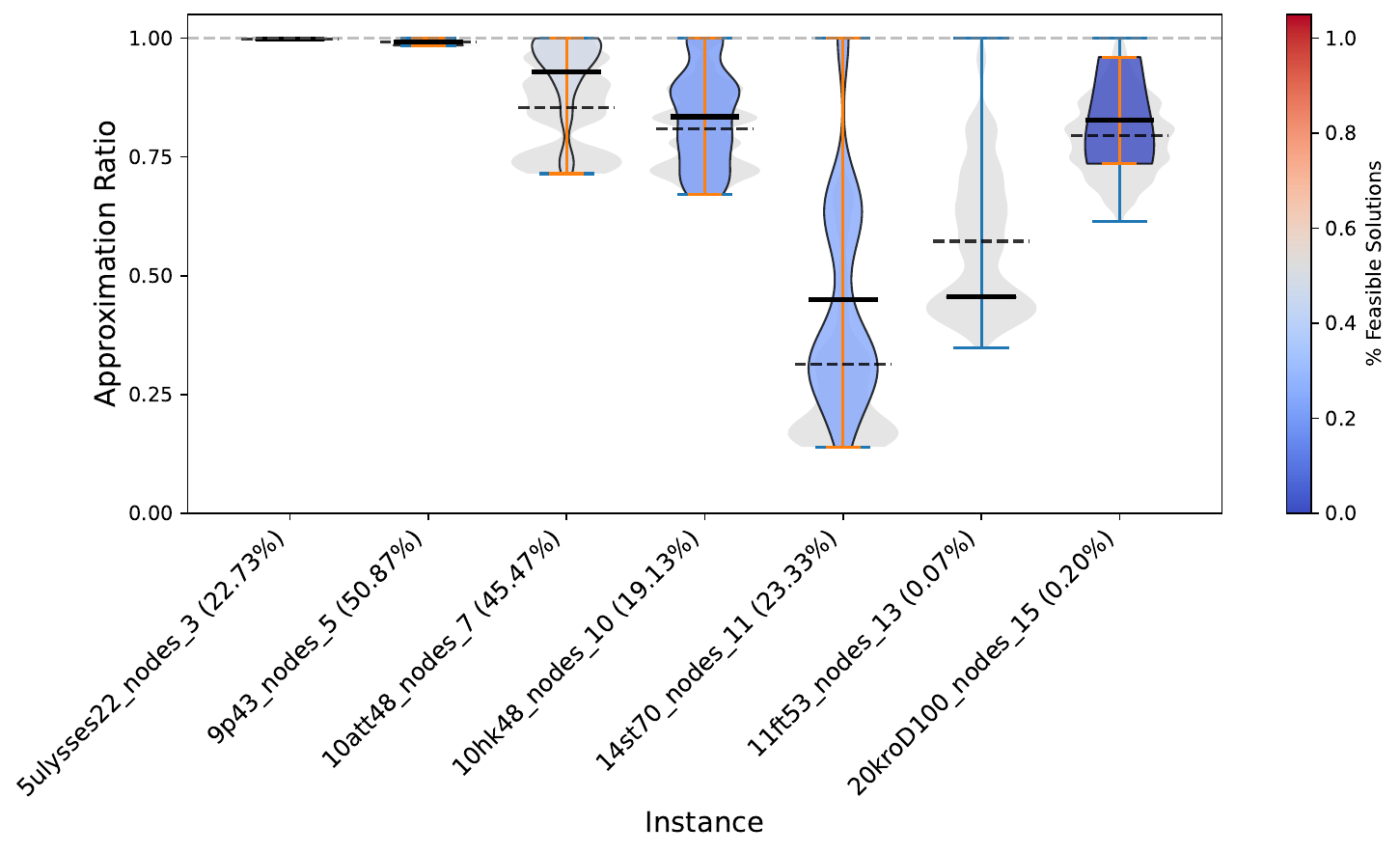}
    \caption{Distribution of solution quality achieved by Q-GTSP on the Subsample Medium dataset with 3 to 20 nodes. The violins show the approximation ratio distribution over sampled feasible solutions, while the instance labels report the feasible-shot rate.}
    \label{fig:subsampling_3-20-QGTSP-approx-violin}
\end{figure}

\begin{figure}
    \centering
    \includegraphics[width=1\columnwidth]{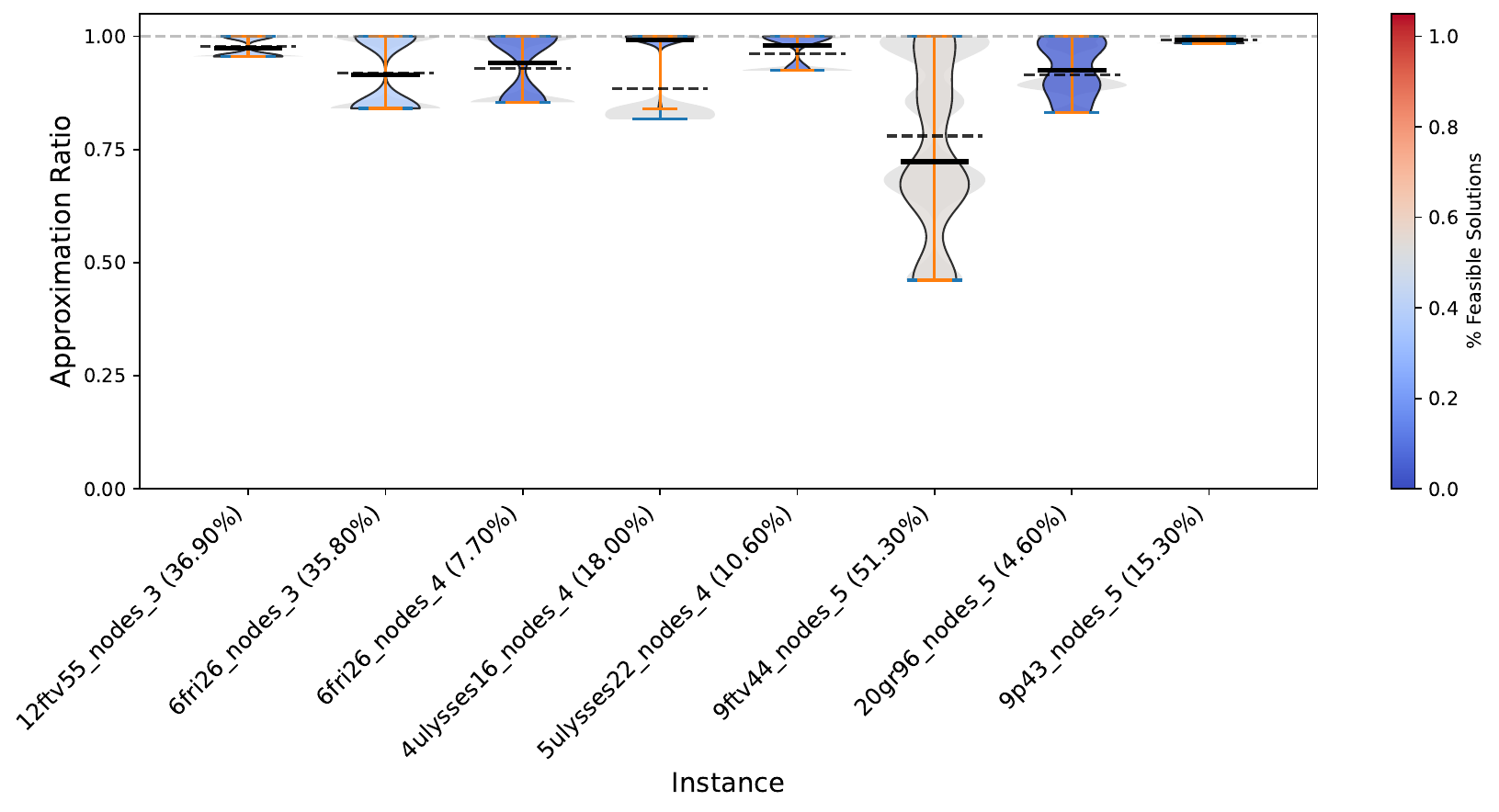}
    \caption{Distribution of solution quality achieved by C-QAOA on the Subsample Small dataset with 3 to 5 nodes. The violins show the approximation ratio distribution over sampled feasible solutions, while the instance labels report the feasible-shot rate.}
    \label{fig:subsampling_3-5-CQAOU-approx-violin}
\end{figure}

\begin{figure}
    \centering
    \includegraphics[width=1\columnwidth]{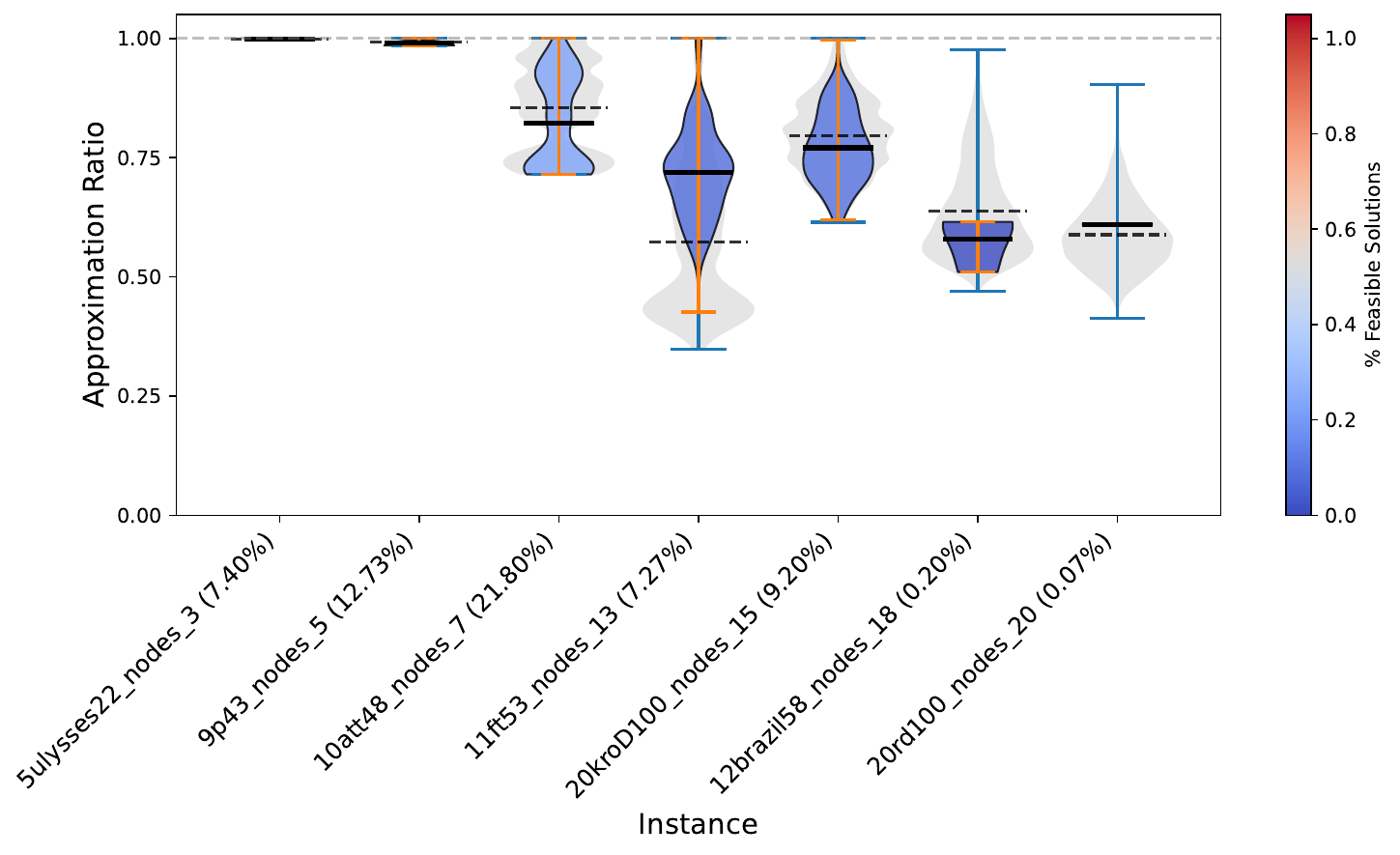}
    \caption{Distribution of solution quality achieved by C-QAOA on the Subsample Medium dataset with 3 to 20 nodes. The violins show the approximation ratio distribution over sampled feasible solutions, while the instance labels report the feasible-shot rate.}
    \label{fig:subsampling_3-20-CQAOU-approx-violin}
\end{figure}

\begin{figure}
    \centering
    \includegraphics[width=1\columnwidth]{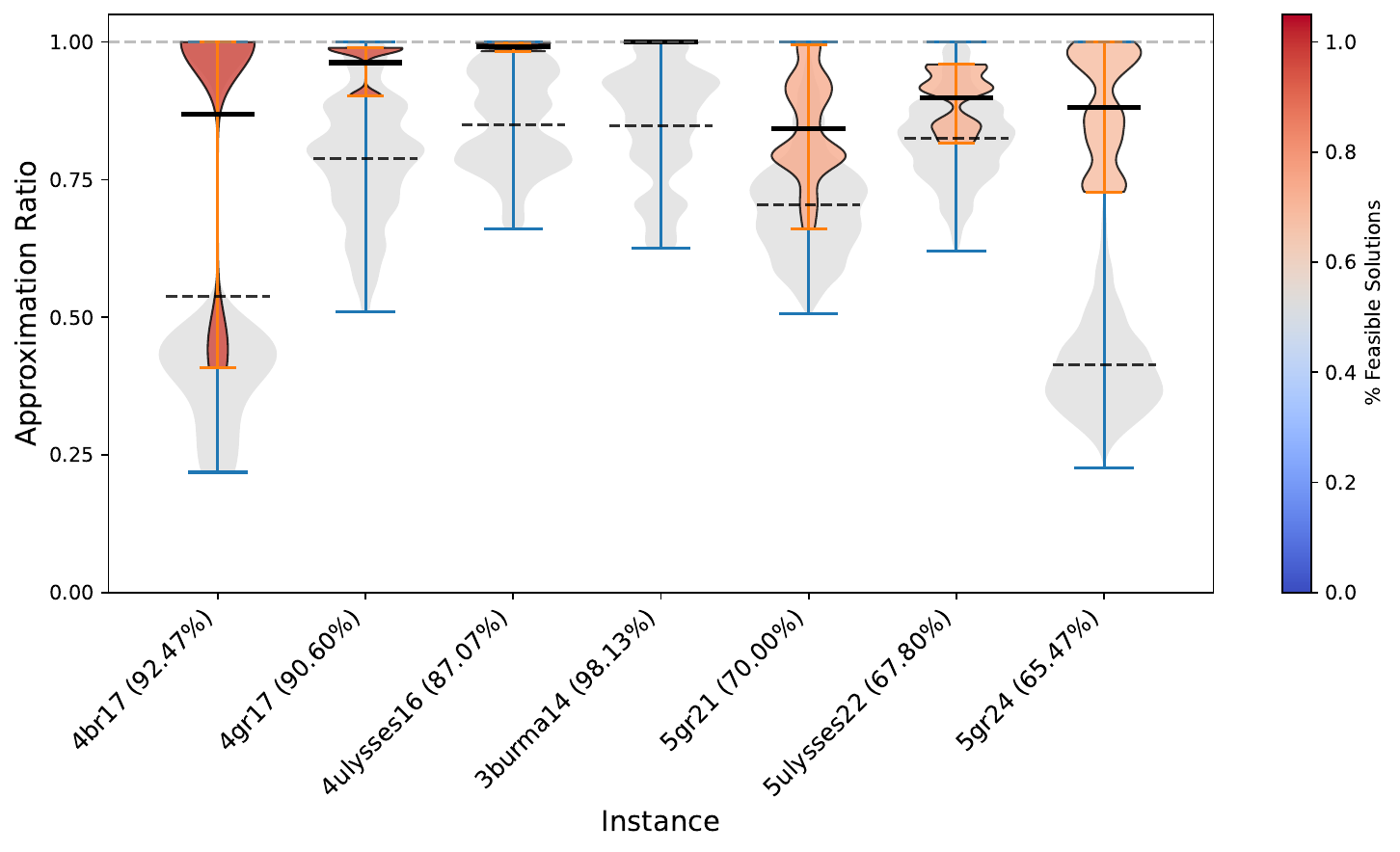}
    \caption{Distribution of solution quality achieved by Q-GTSP on the Preprocess Small dataset (3–5 nodes), where the corresponding original GTSPLIB instances range from 14 to 24 nodes. The violins show the approximation ratio distribution over sampled feasible solutions, while the instance labels report the feasible-shot rate.}
    \label{fig:preprocess_3_5-QGTSP-approx-violin}
\end{figure}

This chapter evaluates the performance of the implemented solvers on the GTSP, comparing solution quality across classical, quantum annealing, and gate-model-based approaches. Our core metrics are the feasibility ratio, i.e., the percentage of solutions that satisfy all given constraints of the optimization problem, and the approximation ratio to measure solution quality, which we define as the quotient between the costs of the optimal solution and the costs of the solution identified by the solver. Similar to the feasibility ratio, an approximation ratio of 1 marks a perfect result, while worse solutions rank at lower approximation ratios, with the worst possible solution ranked at a value that is at least slightly above zero.

\begin{figure*}[t!]
    \centering
    \subfloat[Subsample Small dataset.\label{subfig:ar_subsample_small}]{%
        \includegraphics[width=0.48\linewidth]{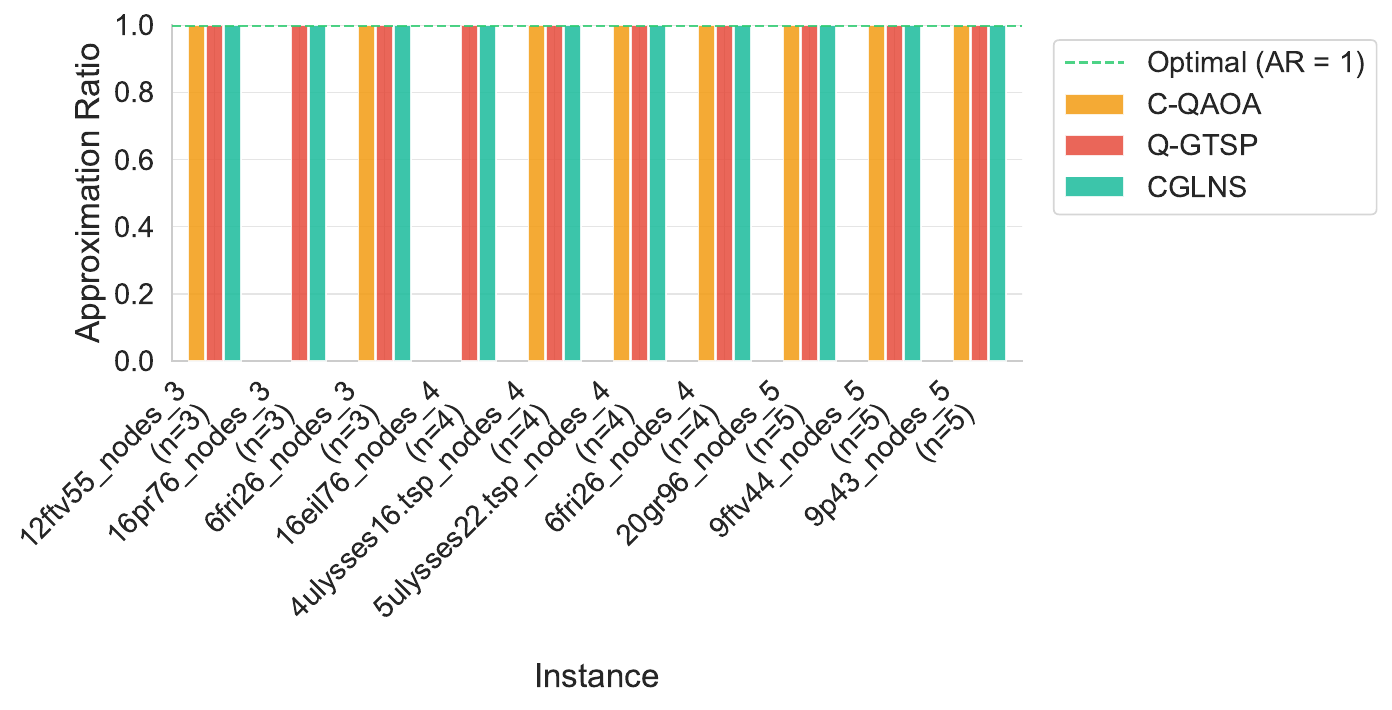}
    }\hfill
    \subfloat[Preprocessing Small dataset.\label{subfig:ar_preprocess_small}]{%
        \includegraphics[width=0.48\linewidth]{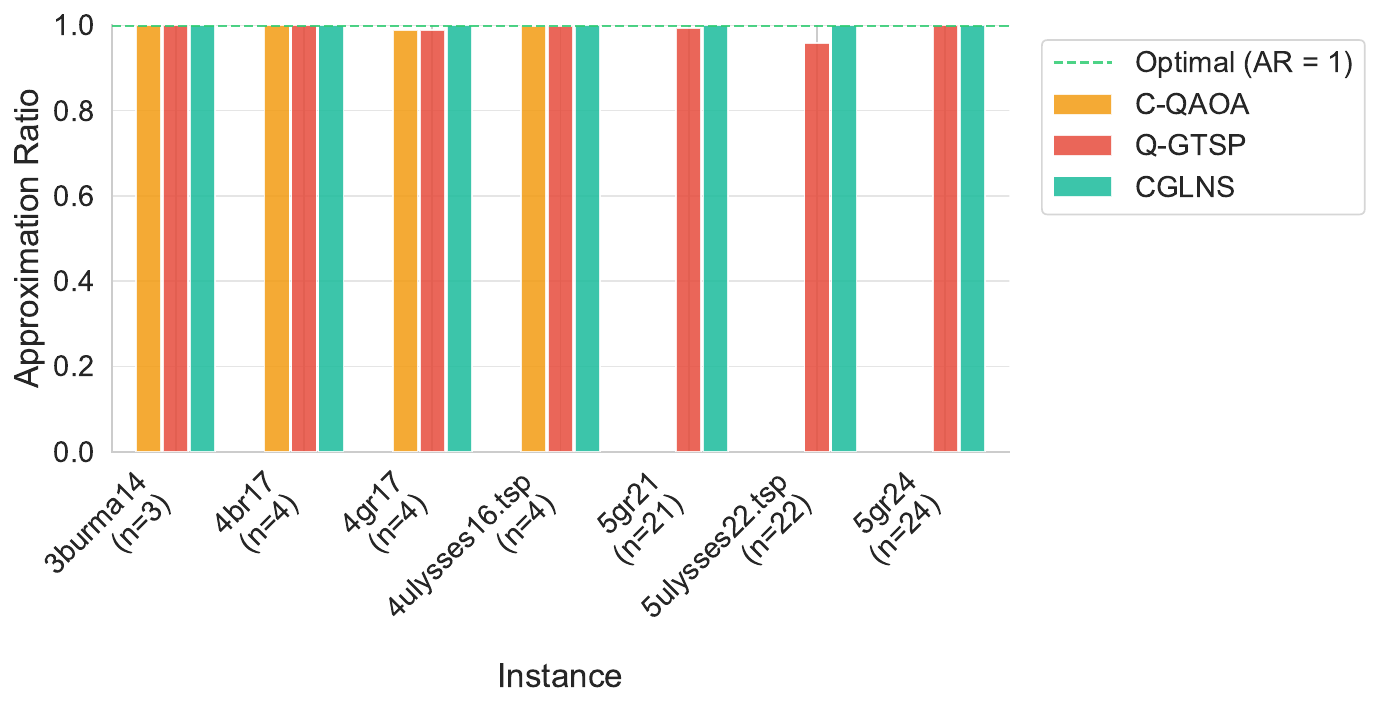}
    }\\[1ex]
    \subfloat[Subsample Medium dataset.\label{subfig:ar_subsample_medium}]{%
        \includegraphics[width=0.48\linewidth]{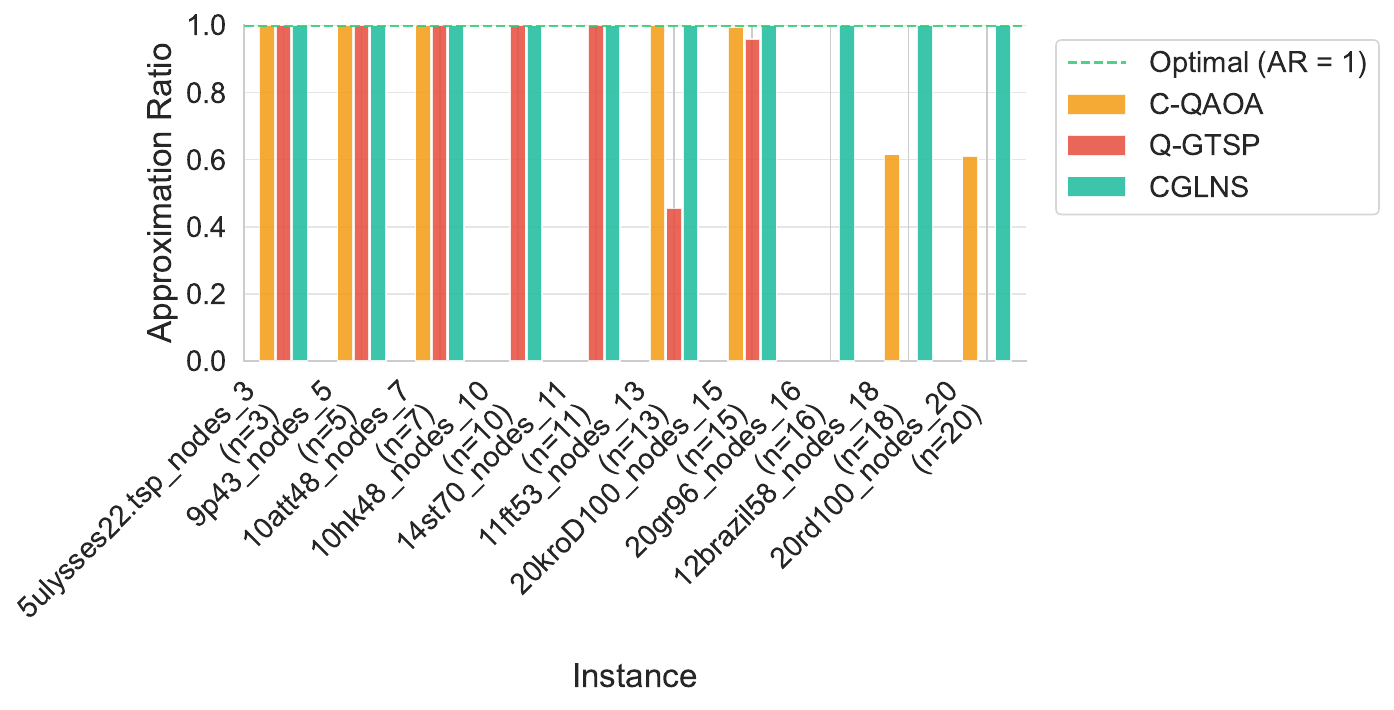}
    }\hfill
    \subfloat[Preprocessing Medium dataset.\label{subfig:ar_preprocess_medium}]{%
        \includegraphics[width=0.48\linewidth]{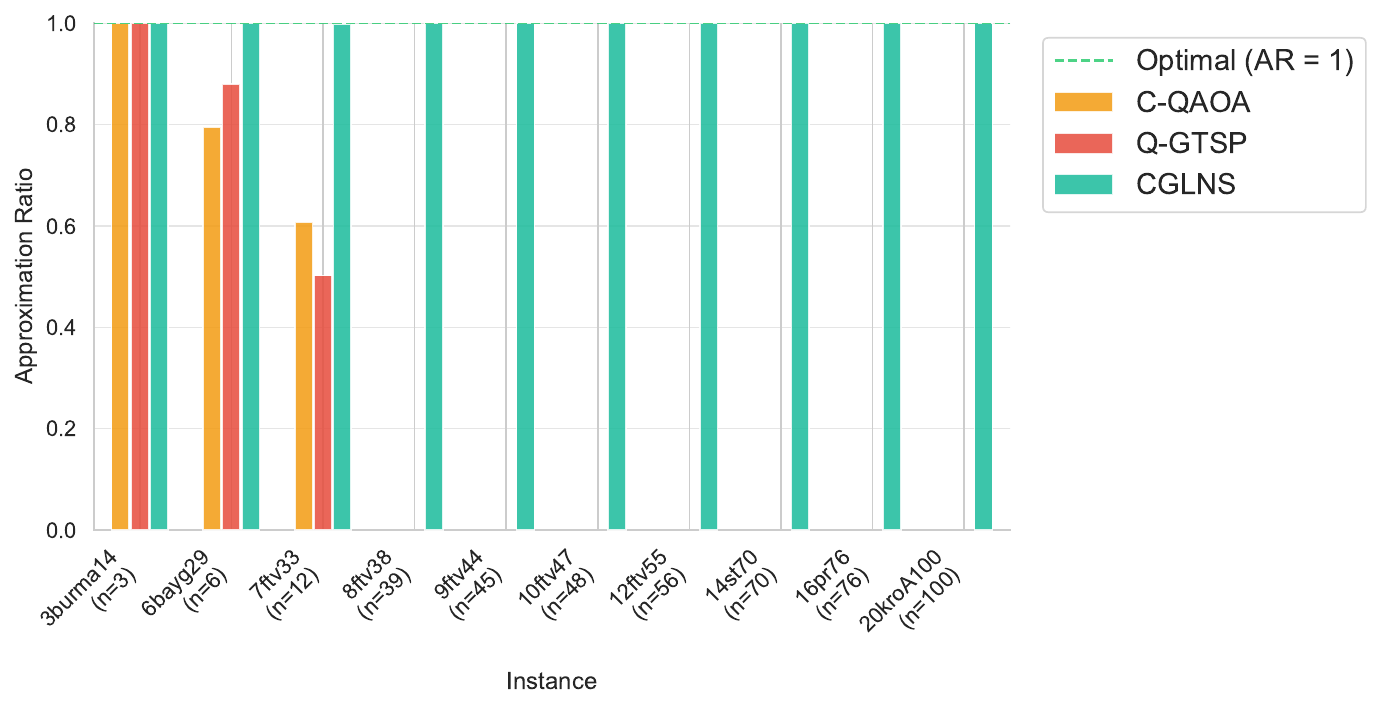}
    }
    \caption{Solution quality of the best shot  measured in approximation ratio at 1500 shots.}
    \label{fig:ar_best_shot}
\end{figure*}

\subsection{Best-Shot Evaluation}
Figure~\ref{fig:ar_best_shot} shows that on most instances, both approaches (QAOA and quantum annealing) consistently find optimal or near optimal solutions until a node count of 13 and above. 
However, expanding the instance sizes to up to 20 nodes decreases the solvers' performance drastically, in particular, the feasibility ratio drops below $50\%$ on all instances, almost reaching $0\%$ on the \texttt{11ft53} instance with 13 nodes. Additionally, the mean is below the random threshold in 4 of the instances, meaning that the majority of sampled solutions are worse than random guessing. Analogous to similar problems, we suspect that the multitude of constraints makes it very hard for the purely QUBO-based quantum annealing approach to find feasible solutions. Similarly, not all constraints are provably satisfied for our QAOA implementation, as the XY-mixers only ensure the satisfaction of the step-wise one-hot constraint but leave open infeasible solutions based on the other two constraints, which are represented by penalty terms instead. For the QAOA in particular, the restriction of $p=1$ further restricts the solution quality for larger problem instances massively. Future work should especially investigate the scaling performance of the quantum solvers based on their time to solution with rising annealing time and QAOA layer depth, respectively.

\subsection{Q-GTSP (Quantum Annealing)}\label{Evaluation:Q-GTSP}

As depicted in Figure~\ref{fig:subsampling_3-5-QGTSP-approx-violin}, the quantum annealing based approach consistently finds optimal solutions on the Subsample Small dataset. For reference, we show the solution distribution over all samples as a color heat map (feasibility) and the landscape of all feasible instances in gray in the background. The bold lines mark the mean solver cost and the dashed lines mark the mean random cost. Notably, the quantum annealer almost always improves the distribution by shifting more probability mass towards the top. 
In some cases however, e.g., the instance \texttt{11ft53} with 13 nodes as shown in Figure~\ref{fig:subsampling_3-20-QGTSP-approx-violin}, we see can see mean of Q-GTSP drop below the mean of for random guessing. This is likely due to a low feasibility ratio, where the few sampled feasible instances are worse than the mean random by chance. 
Considering the feasibility ratio, indicated by the color and the percentage written behind the instance name, we can see that there is a rather large variance in how many feasible solutions are found. Especially for the larger datasets, we can observe a severe drop, as evident in \Cref{fig:subsampling_3-20-QGTSP-approx-violin}. For example, 91\% of the solutions sampled for the instance \texttt{6fri26} are feasible, while only 28\% feasible solutions were found for instance \texttt{16eil76}. This highlights problems regarding the way QUBO deals with constraints. Analogous to other constraint optimization problems, this motivates a quantum solver that can intrinsically reduce the (effective) search space to the space of feasible solutions (cf., e.g., Ref.~\cite{9259965}).

\subsection{C-QAOA}

As depicted in \Cref{fig:subsampling_3-5-CQAOU-approx-violin}, C-QAOA is able to sample optimal or near-optimal tours for most instances of the Subsample Small dataset, as the upper tails of the distributions frequently reach normalized approximation ratio values close to~1. However, the distributions are often broad and partly bimodal. Several instances exhibit probability mass both near the optimum and below the random baseline, indicating that high-quality solutions are reachable but not sampled consistently. Inconsistencies in the solution quality are also partially reflected by the mean values, which remain below the mean of the random baseline for some instances despite the presence of optimal samples. This phenomenon can be explained analogously to the same effect present for quantum annealing (cf. \Cref{Evaluation:Q-GTSP}).

\begin{figure}
    \centering
    \includegraphics[width=0.8\columnwidth]{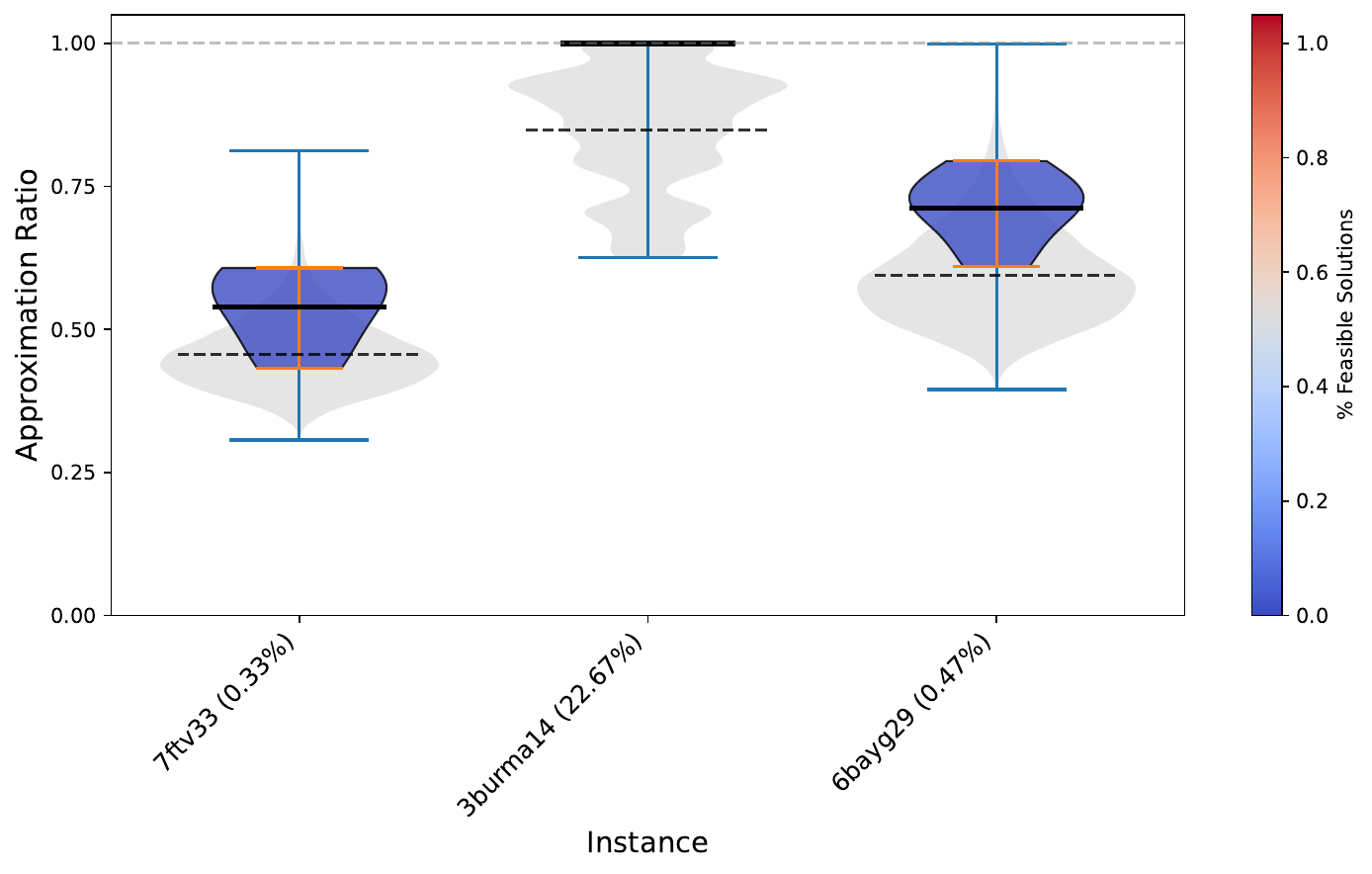}
    \caption{Distribution of solution quality achieved by C-QAOA on the Preprocess Medium dataset (3–20 nodes), where the corresponding original GTSPLIB instances range from 14 to 100 nodes. The violins show the approximation ratio distribution over sampled feasible solutions, while the instance labels report the feasible-shot rate.}
    \label{fig:preprocess_3-20-CQAOU-approx-violin}
\end{figure}

As visualized in \Cref{fig:subsampling_3-20-CQAOU-approx-violin}, the results for the Subsample Medium dataset show a clear degradation in sampling stability. While some instances still contain high-quality samples, the distributions become less concentrated and the mean normalized approximation ratio is more often close to or below the random baseline. In addition, the feasible-shot rate remains low and drops to 0\% for the largest instances. Consequently, C-QAOA frequently fails to provide reliable feasible tours within the fixed shot budget as the instance size increases. An important reason for this phenomenon is likely that the number of utilized QAOA shots (in our case, 1500) is high enough to sample from the complete search space for small problem instances, but it quickly fails to cover the search space when the problem size increases. Overall, Figures~\ref{fig:subsampling_3-5-CQAOU-approx-violin} and~\ref{fig:subsampling_3-20-CQAOU-approx-violin} indicate that C-QAOA can produce competitive solutions on very small GTSP instances up to 5 nodes, but its performance is limited by low and instance-dependent feasibility rates as well as unstable sampling distributions. The main bottleneck is therefore not only the quality of feasible samples, but the probability of observing feasible and high-quality tours within the available number of shots. A similar trend is observed in the preprocessed datasets displayed in \Cref{fig:preprocess_3-20-CQAOU-approx-violin}: while feasible solutions are found for the smaller instances, feasibility degrades rapidly with growing problem size.


\subsection{Comparing Feasibility Ratios}
The results presented so far strongly indicate that the main bottleneck is not only the quality of feasible samples but also the probability of observing feasible tours within the fixed shot budget. Larger instances frequently yield very low feasible-shot rates, as evidenced by Tables~\ref{tab:feasible-subsample-3-20-nodes} and~\ref{tab:feasible-preprocess-3-20-nodes}. 
\begin{table}[htbp]
  \centering
  \caption{Percentage of feasible shots per instance for experiment group \textit{Subsample Medium}. A shot is feasible if it encodes a valid GTSP tour. Cells without shot data show the failure reason.}
  \label{tab:feasible-subsample-3-20-nodes}
  \begin{tabular}{llrr}
    \toprule
    Instance & $n$ & C-QAOA (\%) & QUBO (\%) \\
    \midrule
    5ulysses22\_nodes\_3 & 3 & 7.4 & 22.7 \\
    9p43\_nodes\_5 & 5 & 12.7 & 50.9 \\
    10att48\_nodes\_7 & 7 & 21.8 & 45.5 \\
    10hk48\_nodes\_10 & 10 & Invalid tour & 19.1 \\
    14st70\_nodes\_11 & 11 & Invalid tour & 23.3 \\
    11ft53\_nodes\_13 & 13 & 7.3 & 0.1 \\
    20kroD100\_nodes\_15 & 15 & 9.2 & 0.2 \\
    20gr96\_nodes\_16 & 16 & Invalid tour & Invalid tour \\
    12brazil58\_nodes\_18 & 18 & 0.2 & Invalid tour \\
    20rd100\_nodes\_20 & 20 & 0.1 & Invalid tour \\
    \bottomrule
  \end{tabular}
\end{table}

\begin{table}[htbp]
  \centering
  \caption{Percentage of feasible shots per instance for experiment group \textit{Preprocess Medium}. A shot is feasible if it encodes a valid GTSP tour. Cells without shot data show the failure reason.}
  \label{tab:feasible-preprocess-3-20-nodes}
  \begin{tabular}{llrr}
    \toprule
    Instance & $n$ & C-QAOA (\%) & QUBO (\%) \\
    \midrule
    3burma14 & 3 & 22.7 & 95.4 \\
    6bayg29 & 6 & 0.5 & 26.3 \\
    7ftv33 & 12 & 0.3 & 0.1 \\
    8ftv38 & 13 & Invalid tour & Invalid tour \\
    14st70 & 14 & Invalid tour & Invalid tour \\
    10ftv47 & 15 & Invalid tour & Invalid tour \\
    9ftv44 & 15 & Invalid tour & Invalid tour \\
    16pr76 & 16 & Invalid tour & Could not embed \\
    12ftv55 & 20 & Invalid tour & Could not embed \\
    20kroA100 & 20 & Invalid tour & Could not embed \\
    \bottomrule
  \end{tabular}
\end{table}

Most importantly, quantum annealing performed comparatively better than the QAOA in terms of feasibility ratio. This is likely an artifact of the very low layer depth of the QAOA ($p=1$), in contrast to the standard annealing time available for the quantum annealing based solver. We strongly expect that the QAOA could outperform the quantum annealing solver qualitatively, given more layers and suitable mixers to enforce not just one, but all three constraints.

\subsection{Impact of Preprocessing}
\Cref{fig:preprocess_3_5-QGTSP-approx-violin} indicates that Q-GTSP, consistently achieves approximation ratios that are above the average of the sampled solution landscape for the Preprocess Small instances with 3 to 5 nodes. Across all considered instances, it also shows a high feasibility rate, ranging from approximately 65\% up to 98\%. This is notably higher than the feasibility rates observed for the Subsample Small instances with the same number of nodes (3 to 5), cf. \Cref{fig:subsampling_3-5-QGTSP-approx-violin}.
It should be noted, however, that the preprocessed instances are derived from original GTSPLIB instances and retain the same number of clusters, while only the number of nodes is reduced. Consequently, solutions obtained for preprocessed instances with 3 to 5 nodes correspond to solutions on the original instances, which typically contain between 17 and 24 nodes. This difference should be taken into account when interpreting the results.
In comparison, for the Subsample Medium dataset with 3 to 20 nodes~\ref{fig:subsampling_3-20-QGTSP-approx-violin}, Q-GTSP is only able to find feasible solutions up to instances with 15 nodes, with feasibility rates not exceeding 50\%.
Overall, these observations suggest that preprocessing can improve the ability of Q-GTSP to handle larger underlying problem instances while also increasing the likelihood of obtaining feasible solutions.
Although Q-GTSP with preprocessing still yields approximation ratios above the average of the sampled solution landscape for instances with up to 29 nodes, the feasibility rate decreases noticeably as the instance size increases. One possible explanation for the reduced performance of both solvers on larger preprocessed instances, compared to subsampled instances of similar size, is the higher number of clusters. This structural property, inherited from the original GTSPLIB instances, may increase the problem complexity and negatively affect the solution behavior of the solvers.

In contrast, C-QAOA does not appear to benefit from preprocessing to the same extent. Based on the evaluation, preprocessing can even have a negative impact on C-QAOA performance. While Subsample Small instances of 5 nodes occasionally complete successfully, their Preprocess Small equivalents of the same dimension consistently fail completely, as can be seen in Figure~\ref{fig:preprocess_3-20-CQAOU-approx-violin} and Table~\ref{tab:feasible-preprocess-3-20-nodes}.

More information on the feasibility rates of datasets Subsample Small and Preprocess Small can be found in Appendix~\ref{sec:appendix} (Tables~\ref{tab:feasible-subsample-3-5-nodes} and~\ref{tab:feasible-preprocess-3-5-nodes}). Here we can see that across both small datasets, QUBO consistently achieves higher and more stable feasibility rates than C-QAOA, with preprocessing further increasing QUBO feasibility to up to 98\% while pushing C-QAOA to timeout on all 5-node instances.

\section{Discussion}\label{sec:discussion}

As instance size grows, a consistent decline in solution quality and feasibility can be observed across both quantum approaches. For C-QAOA specifically, feasibility rates drop steeply beyond 5–7 nodes, falling to below 1\% at 13–15 nodes, with timeouts and transpilation failures dominating at larger scales (cf. Tables~\ref{tab:feasible-subsample-3-20-nodes} and~\ref{tab:feasible-preprocess-3-20-nodes}). 
For the quantum annealing based approach, a similar pattern emerges: feasibility rates remain competitive on small instances -- occasionally exceeding 90\% -- but degrade sharply beyond 15 nodes, where infeasibility as well as embedding failures become the dominant issues (cf. Tables~\ref{tab:feasible-subsample-3-20-nodes} and~\ref{tab:feasible-preprocess-3-20-nodes}). Notably, the quantum annealing based solver also encounters embedding failures on preprocessed medium instances, where the number of required logical qubits grows as large as 400, far exceeding current hardware capacity.

Reflecting on the results of both quantum optimization approaches, we find that future quantum solvers must address infeasibility as their main priority. Based on our results, we estimate that the option to address constraints via penalty terms in a QUBO-based approach does not scale in practice. This issue can likely best be addressed by gate-model based approaches such as constraint mixers in the QAOA.

\section{Conclusion}
\label{sec:conclusion}

The Generalized Traveling Salesman Problem arises in a wide range of industrial applications, including logistic routing, scheduling or the optimization of subtask sequences in manufacturing robotics. While there exists substantial prior research on quantum approaches for the TSP, corresponding work on the GTSP is still scarce. In this paper, we introduce a novel method for directly constructing QUBO formulations from GTSP instances, which, to the best of our knowledge, has not been previously described in the literature. We further propose a constrained QAOA solver that expands on this QUBO formulation by implementing one of its one-hot constraints using XY-mixers. We focus on presenting our two quantum GTSP optimizers with current quantum hardware sizes and their limitations in mind.
We evaluate on instances derived from the GTSPLIB benchmark, which we generate via a cluster subset sampling. In addition, we introduced a preprocessing algorithm that reduces the size of GTSPLIB instances before they are passed to the quantum solvers. 
The experimental results show a clear distinction between small and larger instances. For small instances (3 to 5 nodes), the quantum solvers frequently produced solution sets that include the optimal solution or solutions very close to it, indicating performance comparable to classical state-of-the-art solvers. However, as instance size increases, both solution quality and feasibility decrease, and beyond a certain problem size (around 15 nodes and 4 clusters), valid solutions are no longer found reliably.
The proposed preprocessing improves feasibility for the quantum annealing based solver to address larger underlying instances, but its benefits diminish with increasing instance size -- likely due to the higher structural complexity of the preprocessed instances.
Overall, these findings are consistent with results reported in existing literature on quantum optimization, where competitive solution quality is often observed for small instances, but scalability remains a major challenge unless constraints can be properly integrated into the quantum solver.

Future work should focus on improving scalability through proper constraint handling, but also enhanced preprocessing techniques, more efficient encodings, and problem-specific embeddings that better exploit the cluster structure of the GTSP.

\section*{Acknowledgment}
This work has been supported by the LMU Sustainability Fund (EfOiE), the BMFTR (QuCUN, QuaRDS, CAQAO), the Munich Quantum Valley (K5, K7), and the Bavarian StMWi (6GQT). The sole responsibility for the report’s contents lies with the authors.


\appendix
\label{sec:appendix}
\subsection{Additional Metrics}

\begin{table}[H]
  \centering
  \caption{Percentage of feasible shots per instance for experiment group \textit{Subsample Small}. A shot is feasible if it encodes a valid GTSP tour. Cells without shot data show the failure reason.}
  \label{tab:feasible-subsample-3-5-nodes}
  \begin{tabular}{llrr}
    \toprule
    Instance & $n$ & C-QAOA (\%) & QUBO (\%) \\
    \midrule
    12ftv55\_nodes\_3 & 3 & 36.9 & 66.9 \\
    16pr76\_nodes\_3 & 3 & Invalid tour & 41.0 \\
    6fri26\_nodes\_3 & 3 & 35.8 & 48.4 \\
    16eil76\_nodes\_4 & 4 & Invalid tour & 28.7 \\
    4ulysses16\_nodes\_4 & 4 & 18.0 & 38.0 \\
    5ulysses22\_nodes\_4 & 4 & 10.6 & 89.1 \\
    6fri26\_nodes\_4 & 4 & 7.7 & 91.6 \\
    20gr96\_nodes\_5 & 5 & 4.6 & 72.6 \\
    9ftv44\_nodes\_5 & 5 & 51.3 & 88.1 \\
    9p43\_nodes\_5 & 5 & 15.3 & 49.2 \\
    \bottomrule
  \end{tabular}
\end{table}

\begin{table}[H]
  \centering
  \caption{Percentage of feasible shots per instance for experiment group \textit{Preprocess Small}. A shot is feasible if it encodes a valid GTSP tour. Cells without shot data show the failure reason.}
  \label{tab:feasible-preprocess-3-5-nodes}
  \begin{tabular}{llrr}
    \toprule
    Instance & $n$ & C-QAOA (\%) & QUBO (\%) \\
    \midrule
    3burma14 & 3 & 22.5 & 98.1 \\
    4br17 & 4 & 2.6 & 92.5 \\
    4gr17 & 4 & 2.9 & 90.6 \\
    4ulysses16 & 4 & 1.9 & 87.1 \\
    5gr21 & 5 & Timeout & 70.0 \\
    5gr24 & 5 & Timeout & 65.5 \\
    5ulysses22 & 5 & Timeout & 67.8 \\
    \bottomrule
  \end{tabular}
\end{table}

\bibliographystyle{unsrt}  
\bibliography{references} 

@article{Lucas_2014_Ising_formulations,
   title={Ising formulations of many NP problems},
   volume={2},
   ISSN={2296-424X},
   url={http://dx.doi.org/10.3389/fphy.2014.00005},
   DOI={10.3389/fphy.2014.00005},
   journal={Frontiers in Physics},
   publisher={Frontiers Media SA},
   author={Lucas, Andrew},
   year={2014} }

@article{Born1928,
    author = {Born, M and Fock, V},
    doi = {10.1007/BF01343193},
    issn = {0044-3328},
    journal = {Zeitschrift f{\"{u}}r Phys.},
    number = {3},
    pages = {165--180},
    title = {{Beweis des Adiabatensatzes}},
    url = {https://doi.org/10.1007/BF01343193},
    volume = {51},
    year = {1928}
}

@article{SMITH20171,
title = {GLNS: An effective large neighborhood search heuristic for the Generalized Traveling Salesman Problem},
journal = {Computers \& Operations Research},
volume = {87},
pages = {1-19},
year = {2017},
issn = {0305-0548},
doi = {https://doi.org/10.1016/j.cor.2017.05.010},
author = {Stephen L. Smith and Frank Imeson},
}

@INPROCEEDINGS{9259965,
  author={Bärtschi, Andreas and Eidenbenz, Stephan},
  booktitle={2020 IEEE International Conference on Quantum Computing and Engineering (QCE)}, 
  title={Grover Mixers for QAOA: Shifting Complexity from Mixer Design to State Preparation}, 
  year={2020},
  volume={},
  number={},
  pages={72-82},
  doi={10.1109/QCE49297.2020.00020}}

@misc{glover2019tutorialformulatingusingqubo,
      title={A Tutorial on Formulating and Using QUBO Models}, 
      author={Fred Glover and Gary Kochenberger and Yu Du},
      year={2019},
      eprint={1811.11538},
      archivePrefix={arXiv},
      primaryClass={cs.DS},
      url={https://arxiv.org/abs/1811.11538}, 
}

@article{GTSP_Def,
title = {A comprehensive survey on the generalized traveling salesman problem},
journal = {European Journal of Operational Research},
volume = {314},
number = {3},
pages = {819-835},
year = {2024},
issn = {0377-2217},
doi = {https://doi.org/10.1016/j.ejor.2023.07.022},
url = {https://www.sciencedirect.com/science/article/pii/S0377221723005581},
author = {Petrică C. Pop and Ovidiu Cosma and Cosmin Sabo and Corina Pop Sitar},
}

@InProceedings{10.1007/978-981-97-7801-0_12,
author="He, Haoqi",
editor="Ghosh, Smita
and Zhang, Zhao",
title="Quantum Annealing and GNN for Solving TSP with QUBO",
booktitle="Algorithmic Aspects in Information and Management",
year="2024",
publisher="Springer Nature Singapore",
address="Singapore",
}

@book{nielsen_chuang_2010,
    place={Cambridge},
    title={Quantum Computation and Quantum Information: 10th Anniversary Edition}, DOI={10.1017/CBO9780511976667},
    publisher={Cambridge University Press},
    author={Nielsen, Michael A. and Chuang, Isaac L.},
    year={2010}
}

@misc{GTSPLIB,
  author = {Zverovich, Alexei},
  title = {{GTSPLIB} -- GTSP Instances Library},
  year = {2002},
  howpublished = {\url{https://www.cs.rhul.ac.uk/home/zvero/GTSPLIB/}},
  note = {Accessed: 2026-01-26}
}

@article{el2021pre,
  title={A pre-processing reduction method for the generalized travelling salesman problem},
  author={El Krari, Mehdi and Ahiod, Bela{\"\i}d and El Benani, Youssef Bouazza},
  journal={Operational Research},
  volume={21},
  number={4},
  pages={2543--2591},
  year={2021},
  publisher={Springer}
}

@misc{dwave_quantum_annealing_intro,
  title        = {What is Quantum Annealing?},
  author       = {{D‑Wave Quantum}},
  year         = 2026,
  url          = {https://docs.dwavequantum.com/en/latest/quantum_research/quantum_annealing_intro.html},
  note         = {Accessed: 2026-01-27}
}

@inproceedings{nusslein2022algorithmic,
  title={Algorithmic QUBO formulations for k-SAT and hamiltonian cycles},
  author={N{\"u}{\ss}lein, Jonas and Gabor, Thomas and Linnhoff-Popien, Claudia and Feld, Sebastian},
  booktitle={Proceedings of the genetic and evolutionary computation conference companion},
  pages={2240--2246},
  year={2022}
}

@misc{dwave_advantage_quantum_computer_2026,
  title        = {The Advantage™ Quantum Computer},
  author       = {{D‑Wave Quantum Inc.}},
  year         = {2026},
  howpublished = {\url{https://www.dwavequantum.com/solutions-and-products/systems/}},
  note         = {Accessed: 2026‑01‑30},
}

@misc{ibm_quantum_processor_types_2026,
  title        = {Processor types},
  author       = {{IBM Quantum Documentation}},
  year         = {2026},
  howpublished = {\url{https://quantum.cloud.ibm.com/docs/en/guides/processor-types}},
  note         = {Accessed: 2026-01-30},
}

@article{chander2024solving,
  title={Solving the traveling salesperson problem with quantum approximate optimization algorithms},
  author={Chander, Siddarth and Sathishkumar, Naren and Hussain, Affan and Blekos, Kostas},
  journal={Preprint},
  year={2024}
}

@article{silva2021mapping,
  title={Mapping a logical representation of TSP to quantum annealing.},
  author={Silva, Carla and Aguiar, Ana and Lima, Priscila and Dutra, In{\^e}s},
  journal={Quantum Information Processing},
  volume={20},
  number={12},
  year={2021}
}

@article{qian2023comparative,
  title={Comparative study of variations in quantum approximate optimization algorithms for the traveling salesman problem},
  author={Qian, Wenyang and Basili, Robert AM and Eshaghian-Wilner, Mary Mehrnoosh and Khokhar, Ashfaq and Luecke, Glenn and Vary, James P},
  journal={Entropy},
  volume={25},
  number={8},
  pages={1238},
  year={2023},
  publisher={MDPI}
}

@article{padmasola2025solving,
  title={Solving the Traveling Salesman Problem via Different Quantum Computing Architectures},
  author={Padmasola, Venkat and Li, Zhaotong and Chatterjee, Rupak and Dyk, Wesley},
  journal={arXiv preprint arXiv:2502.17725},
  year={2025}
}

@article{kalleri2025edge,
  title={An edge-based and subspace reduction encoding scheme to solve the traveling salesman problem in quantum computers},
  author={Kalleri Madhu, Anandu and Li, Chi-Kwong and R{\"o}nkk{\"o}, Jami and Nakahara, Mikio and Lee, Ray-Kuang},
  journal={arXiv e-prints},
  pages={arXiv--2512},
  year={2025}
}

@article{bun2020controlling,
  title={Controlling privacy loss in sampling schemes: An analysis of stratified and cluster sampling},
  author={Bun, Mark and Drechsler, J{\"o}rg and Gaboardi, Marco and McMillan, Audra and Sarathy, Jayshree},
  journal={arXiv preprint arXiv:2007.12674},
  year={2020}
}

@article{farhi2014quantum,
  title={A quantum approximate optimization algorithm},
  author={Farhi, Edward and Goldstone, Jeffrey and Gutmann, Sam},
  journal={arXiv preprint arXiv:1411.4028},
  year={2014}
}

@article{hadfield2019quantum,
  title={From the quantum approximate optimization algorithm to a quantum alternating operator ansatz},
  author={Hadfield, Stuart and Wang, Zhihui and O’gorman, Bryan and Rieffel, Eleanor G and Venturelli, Davide and Biswas, Rupak},
  journal={Algorithms},
  volume={12},
  number={2},
  pages={34},
  year={2019},
  publisher={MDPI}
}

@inproceedings{Dhami_2023,
   title={GATSBI: An Online GTSP-Based Algorithm for Targeted Surface Bridge Inspection},
   url={http://dx.doi.org/10.1109/ICUAS57906.2023.10156013},
   DOI={10.1109/icuas57906.2023.10156013},
   booktitle={2023 International Conference on Unmanned Aircraft Systems (ICUAS)},
   publisher={IEEE},
   author={Dhami, Harnaik and Yu, Kevin and Williams, Troi and Vajipey, Vineeth and Tokekar, Pratap},
   year={2023},
   month=jun, pages={1199–1206} }

@misc{suarezruiz2017robotspfastsolution,
      title={RoboTSP - A Fast Solution to the Robotic Task Sequencing Problem}, 
      author={Francisco Suárez-Ruiz and Teguh Santoso Lembono and Quang-Cuong Pham},
      year={2017},
      eprint={1709.09343},
      archivePrefix={arXiv},
      primaryClass={cs.RO},
      url={https://arxiv.org/abs/1709.09343}, 
}

@article{bako2025prog,
  title={Prog-QAOA: Framework for resource-efficient quantum optimization through classical programs},
  author={Bak{\'o}, Bence and Glos, Adam and Salehi, {\"O}zlem and Zimbor{\'a}s, Zolt{\'a}n},
  journal={Quantum},
  volume={9},
  pages={1663},
  year={2025},
  publisher={Verein zur F{\"o}rderung des Open Access Publizierens in den Quantenwissenschaften}
}

@article{glos2022space,
  title={Space-efficient binary optimization for variational quantum computing},
  author={Glos, Adam and Krawiec, Aleksandra and Zimbor{\'a}s, Zolt{\'a}n},
  journal={npj Quantum Information},
  volume={8},
  number={1},
  pages={39},
  year={2022},
  publisher={Nature Publishing Group UK London}
}

@article{farhi2000quantum,
  title={Quantum computation by adiabatic evolution},
  author={Farhi, Edward and Goldstone, Jeffrey and Gutmann, Sam and Sipser, Michael},
  journal={arXiv preprint quant-ph/0001106},
  year={2000}
}

@article{sack2021quantum,
  title={Quantum annealing initialization of the quantum approximate optimization algorithm},
  author={Sack, Stefan H and Serbyn, Maksym},
  journal={quantum},
  volume={5},
  pages={491},
  year={2021},
  publisher={Verein zur F{\"o}rderung des Open Access Publizierens in den Quantenwissenschaften}
}

@article{shirai2025compressed,
  title={Compressed space quantum approximate optimization algorithm for constrained combinatorial optimization},
  author={Shirai, Tatsuhiko and Togawa, Nozomu},
  journal={IEEE Transactions on Quantum Engineering},
  year={2025},
  publisher={IEEE}
}

@article{bucher2025penalty,
  title={Penalty-free approach to accelerating constrained quantum optimization},
  author={Bucher, David and Stein, Jonas and Feld, Sebastian and Linnhoff-Popien, Claudia},
  journal={Physical Review A},
  volume={112},
  number={6},
  pages={062605},
  year={2025},
  publisher={APS}
}

@article{karapetyan2012efficient,
  title={Efficient local search algorithms for known and new neighborhoods for the generalized traveling salesman problem},
  author={Karapetyan, Daniel and Gutin, Gregory},
  journal={European Journal of Operational Research},
  volume={219},
  number={2},
  pages={234--251},
  year={2012},
  publisher={Elsevier}
}

@article{laporte1996some,
  title={Some applications of the generalized travelling salesman problem},
  author={Laporte, Gilbert and Asef-Vaziri, Ardavan and Sriskandarajah, Chelliah},
  journal={Journal of the Operational Research Society},
  volume={47},
  number={12},
  pages={1461--1467},
  year={1996},
  publisher={Taylor \& Francis}
}

@article{baniasadi2020transformation,
  title={A transformation technique for the clustered generalized traveling salesman problem with applications to logistics},
  author={Baniasadi, Pouya and Foumani, Mehdi and Smith-Miles, Kate and Ejov, Vladimir},
  journal={European Journal of Operational Research},
  volume={285},
  number={2},
  pages={444--457},
  year={2020},
  publisher={Elsevier}
}

@article{abbas2024challenges,
  title={Challenges and opportunities in quantum optimization},
  author={Abbas, Amira and Ambainis, Andris and Augustino, Brandon and B{\"a}rtschi, Andreas and Buhrman, Harry and Coffrin, Carleton and Cortiana, Giorgio and Dunjko, Vedran and Egger, Daniel J and Elmegreen, Bruce G and others},
  journal={Nature Reviews Physics},
  volume={6},
  number={12},
  pages={718--735},
  year={2024},
  publisher={Nature Publishing Group}
}

@misc{gurobi,
  author = {{Gurobi Optimization, LLC}},
  title = {{Gurobi Optimizer Reference Manual}},
  year = 2026,
  url = "https://www.gurobi.com"
}

\end{document}